\documentclass[aip,graphicx]{revtex4-1}
\usepackage{graphics,amsmath,amssymb,graphicx}
\usepackage{xcolor}
\draft 

\begin{document}

\title{TNSA based proton acceleration by two oblique laser pulses in the presence of an axial magnetic field} 

\author{Imran Khan}
\email[]{imran.khan@physics.iitd.ac.in}

\affiliation{Department of Physics, Indian Institute of  Technology Delhi, Hauz Khas, New Delhi, India-110016}

\author{Vikrant Saxena}
\email[]{vsaxena@physics.iitd.ac.in}
\affiliation{Department of Physics, Indian Institute of  Technology Delhi, Hauz Khas, New Delhi, India-110016}

\date{\today}
\begin{abstract}
A recently proposed strategy to boost the proton/ion cutoff energy in the target normal sheath acceleration scheme employs two obliquely incident laser pulses simultaneously irradiating the flat target rather than a single normally incident laser pulse of twice the pulse energy. Moreover, the presence of an externally applied magnetic field along the normal of the target's rear surface is known to reduce the angular divergence of hot electrons which results in a more efficient sheath field at the target rear leading to increased cutoff energy of accelerated protons/ions. In the present work, we employ two-dimensional Particle-In-Cell (PIC) simulations to examine, in detail, the effect of such a magnetic field on the cutoff energy of protons/ions in the cases of normal as well as oblique incidence of the laser pulse on a flat target. It is shown that the two-oblique-pulse configuration combined with an external magnetic field results in a stronger enhancement of the cutoff energies as compared to the normal incidence case.

\end{abstract}

\maketitle 
\section{Introduction}

The acceleration of protons/ions via irradiation of solid targets with high-intensity laser pulses has attracted wide research interest over the last couple of decades. This has been largely motivated by the possibility of significantly reduced dimensions of the future accelerators owing to the strong electric fields ($\approx 10^{13}$ V/m) that are created during this highly complex interaction process. Due to the compact dimensions involved, it is expected to be a more economical option for potential applications, including isochoric heating\cite{patel2003isochoric}, fast ignition\cite{roth2001fast, atzeni2002first}, examining short-lived electric and magnetic fields in plasma\cite{borghesi2002electric, borghesi2003measurement}, hadron therapy\cite{bulanov2002feasibility, ledingham2014towards}, and so on.
However, certain challenges are yet to be overcome before the demands of most of these applications are met. The paramount issues have been to enhance the cut-off energy of the accelerated protons/ions while reducing their divergence.


There are several acceleration mechanisms depending upon the laser pulse and target parameters, namely, target normal sheath acceleration (TNSA), radiation pressure acceleration (RPA), etc. Due to the moderate laser pulse requirements the TNSA \cite{wilks2001energetic, snavely2000intense, mora2003plasma} mechanism of proton/ion acceleration has been widely investigated theoretically as well as experimentally, however, the ion/proton beam accelerated by the TNSA mechanism has exponential energy spectra and only a small fraction of them are highly energetic. Moreover, in TNSA the cutoff energy of the accelerated ions/protons is constrained by laser energy according to, $E_{max}\ \propto \  I^\beta$, with $\beta <1 $, where $\beta$ depends on the laser pulse duration.

It is therefore essential to examine approaches to improve the number of energetic ions/protons and enhance their cutoff energies. Recently, Ferri et al.\cite{ferri2019enhanced} have reported that when a single laser pulse is split into two identical laser pulses of half the intensity (without increasing the overall laser energy) which irradiate the target obliquely at two different angles, there is an improvement both in the number of energetic ions/protons and in their cutoff energy. This was further investigated by Nashad et al.\cite{rahman2021particle} to understand the effect of phase difference, time delay, and spatial separation between the two pulses on the proton energy spectra.

Another challenge in the laser-based ion/proton acceleration mechanisms remains to meet energy spread and the divergence requirement of many applications, e.g., hadron treatment of cancer cells, where the energy spread of the proton beam should be exceedingly low, ideally $1\%$ or below\cite{daido2012review}, and low beam divergence is also necessary to protect the healthy cells.

A strategy to reduce the divergence of energetic electrons and thus of accelerated protons/ions is to introduce a longitudinal magnetic field of kilo-Tesla level. Thanks to improvements in the generation of magnetic fields by laser-driven coil targets that vary from hundreds of Tesla to kilo-Tesla\cite{fujioka2013kilotesla, santos2015laser, gao2016ultrafast, kim2016radiation}. The GEKKO-XII\cite{fujioka2013kilotesla} laser facility has been reported to produce a magnetic field of 1.5 kT across a distance of several hundred microns. This field degrades at a rate of 3 T/ps and for the femtosecond laser-plasma interaction, it can be considered as constant. 
The externally applied magnetic field of the kilo-Tesla level restricts the transverse motion of the hot electrons and guides them along the longitudinal direction. This produces a better-focused electron stream, which leads to the creation of a stronger sheath and, as a result, a higher cutoff energy for ions and protons. Such an enhancement of energy cut off of protons has been reported by considering a normal incidence of the laser pulse on a flat target in the presence of kilo-Tesla level magnetic field\cite{arefiev2016enhanced, weichman2020generation}.\\

In this paper, we examine the effect of oblique incidence of laser pulses on the flat target in the presence of an external longitudinal magnetic field by employing 2D particle-in-cell(PIC) simulations. We first investigate in detail the case of a normally incident laser pulse to understand the acceleration mechanism with and without the external magnetic field. We then extend our investigations to the case of oblique incidence of the laser pulse in the presence of a kilo-Tesla level external magnetic field along the normal to the target rear surface. In particular, we consider two obliquely incident laser pulses of half the intensity of the single laser pulse used in the normal incidence case, as proposed by Ferri et al\cite{ferri2019enhanced}, and examine the effect of the external magnetic field. Finally, the case of two obliquely incident laser pulses is compared with the case of a single normally incident laser pulse. For completeness, we also report results for the case of a single obliquely incident laser pulse in the presence of an external magnetic field.
 
\begin{figure}
	\includegraphics[height=4cm,width=0.4\textwidth]{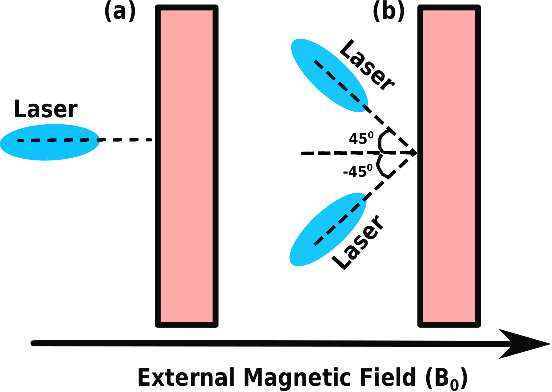}
 	\caption{\label{fig:epsart} Schematic representation of normal(left) and oblique (right) incident laser pulse onto the flat target in the presence of an externally applied longitudinal magnetic field.}
  \label{fig:schematic}
\end{figure}

In the following section, we provide the details of our simulations. In section III, the simulation results are discussed for the case of normal incidence of the laser pulse with and without an external magnetic field. In section IV, the case of two oblique laser pulses is presented and the results are compared with the normal incidence case. In the final section, the results are concluded and future directions are envisaged. 

\section{Simulation Setup}
Two-dimensional (2D) particle-in-cell (PIC) simulations are carried out with open-source PIC code EPOCH\cite{arber2015contemporary}. The simulation box is 90 $\mu$m $\times$ 189 $\mu$m with 10000 $\times$ 7000 cells along the x and y-axis, respectively. The simulation box extends from -10 $\mu$m to 80 $\mu$m along the x-axis i.e. along the direction of laser propagation, and $ \pm$94.5 $\mu$m along the y-axis. In the transverse direction of the simulation box, thermal and periodic boundaries are used for particles and fields, respectively, while open and simple laser boundaries are used at the right and left ends.

The target is localized between 0 to 7 $\mu$m along the x-axis and between $\pm$ 94 $\mu$m along the y-axis and is composed of fully ionized polyethylene [(C$_2$H$_4$)$_n$]. The mass density of polyethylene is $\rho$ = 0.93 g/cm$^3 $ which corresponds to the number density of carbon ions = 22.88 $n_c$, protons = 45.76 $n_c$, and electrons = 183.0 6$n_c$. Where $n_c$ = $\epsilon _{\circ} m_e\omega_{\circ}^2 /e^2$ is the critical density. For carbon ions, 20 macroparticles per cell, while for protons as well as for electrons, 60 macroparticles per cell have been used. 

The p-polarized laser pulse used in our simulations has a wavelength of 0.8 $\mu$m and an intensity of 5.5$\times 10^{20}$  W/cm $^{2}$. The laser pulse is propagating along the +ve x-axis and has a Gaussian profile both in space and time. The focal spot of the laser pulse at the waist is $3\mu$m and the duration(FWHM) is 40 fs. These laser parameters are similar to those used in Ref.\cite{scullion2017polarization} and are comparable to the experimental set-up at Rutherford Appleton Lab (RAL), STFC, UK. Along with the laser, a constant, uniform magnetic field ($B_0 = 2kT$) is applied along the x-axis. This field serves as an externally imposed, quasi-static magnetic field in the experiment.

We consider two scenarios, in the first one(SNP), a single laser pulse is incident normally at the target front surface. In the second scenario (TCP), the laser pulse considered for the SNP configuration is split into two oblique pulses of equal energy, and half the intensity. These pulses are incident onto the target from two different directions ($\pm$45$^{\circ}$). Dashed and solid lines are used to represent the with and without external magnetic field cases, respectively, in both the above scenarios. \\

\begin{figure}
	\includegraphics[height=4.5cm,width=1\textwidth]{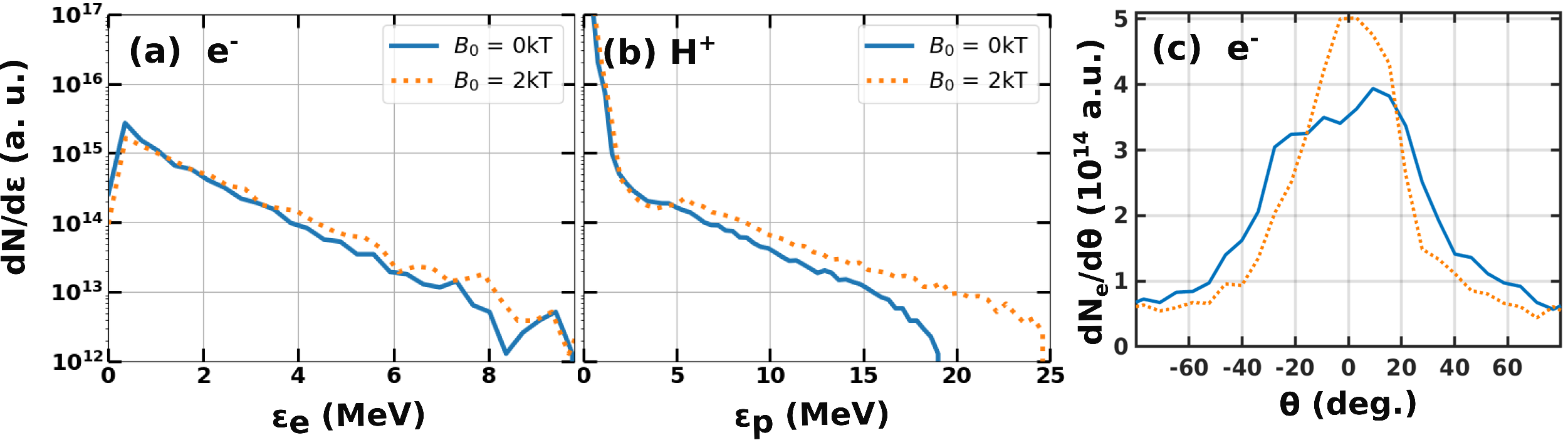}
 	\caption{ (a) Electron energy distribution, (b) protons energy distribution, and (c) electron angular distribution without (solid line) and with (dotted line) an external magnetic field. The electron energy distribution and angular distributions are shown at t=100 fs whereas the energy distribution of the proton is recorded at t=870 fs. }
  \label{fig:energy_distribution}
\end{figure}

\section{Normal incidence in the presence of longitudinal B}
In the TNSA mechanism, the laser pulse interacts with a solid target and transfers a significant part of its energy to the electrons. These hot electrons exit at the rear side without losing much energy since their collision mean-free path is much longer than the target thickness. The ions and protons are accelerated by the strong sheath field that is formed at the target rear by the hot electrons. In Fig. \ref{fig:energy_distribution}, the energy spectra of electrons (left), protons (middle), and angular distribution of electrons (right) are shown for the single normal pulse (SNP) with and without the magnetic field.

\begin{figure}
	\includegraphics[width=.5\textwidth]{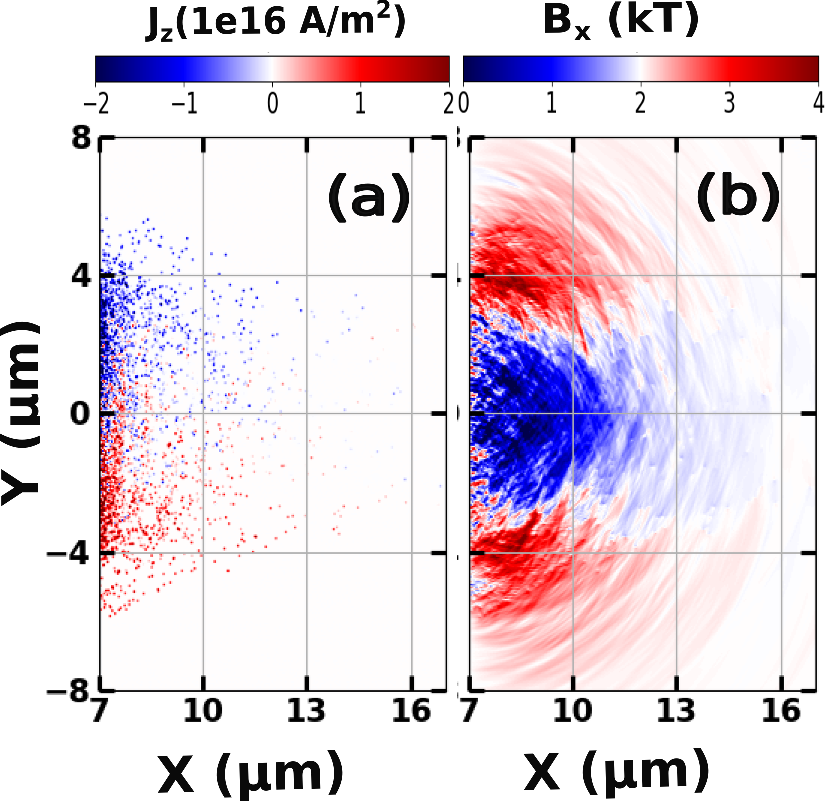}
 	\caption{ Induced current density along the z-direction, $j_z$ (left), and modified longitudinal magnetic field (right) at the rear side of the target, at time t=100 fs, in the presence of an external magnetic field.}
  \label{fig:jz_bx}
\end{figure}
Figure \ref {fig:energy_distribution}a, shows that the externally applied magnetic field has a negligible effect on the electron energy spectra but the protons show sufficient enhancement in their cutoff energy (Fig. \ref {fig:energy_distribution}b). The applied magnetic field ($B_0$ = 2kT) and electron cyclotron frequency (~3.5e14 Hz) are much lower than that of the laser magnetic field (~2e5T) and frequency (~23.5e14 Hz), which makes it impossible for resonant heating. The angular distribution of electrons in Fig. \ref{fig:energy_distribution}c shows that the applied magnetic field restricts electrons' motion in the transverse direction to some extent, and focuses them along the longitudinal direction.

\begin{figure}
	\includegraphics[width=1\textwidth]{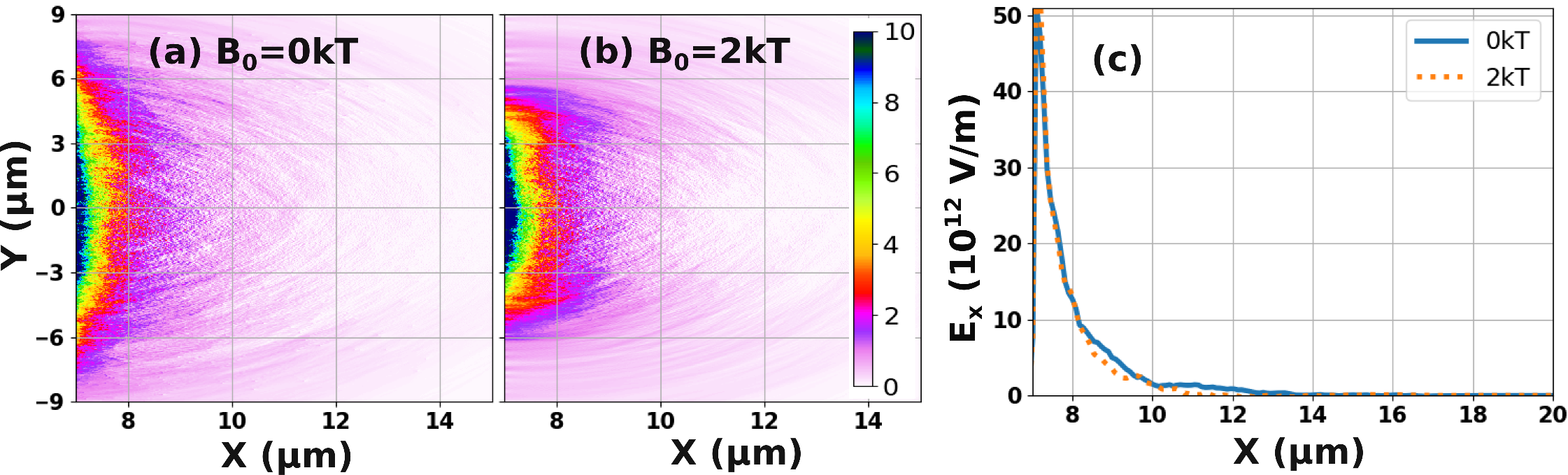}
 	\caption{The longitudinal electric field at the rear side of the target without (a), with (b) an external magnetic field and (c) lineout of these two at time t=100 fs.}
  \label{fig:sheath}
\end{figure}

The transverse electric field is associated with the expanding plasma plume at the rear side and points outward on both sides. The longitudinal magnetic field combined with this electric field gives rise to the $\bold{E\times B_0}$ current in the z-direction (Fig. \ref{fig:jz_bx}a). This current exists as the number of electrons at the rear side of the target is much higher than the positive charges. This current points into the plain and out of the plain of paper, above and below the x-axis, respectively, and induces its own magnetic field ($\bold{B_i}$). The induced magnetic field is clockwise and anticlockwise above and below the x-axis. The resulting magnetic field inside is ($\bold{B_0 - B_i}$) and outside the plasma plume is ($\bold{B_0 + B_i}$). Hence the generated current in the z-direction results in the longitudinal magnetic field reduced inside and enhanced outside the plasma as shown in Fig. \ref{fig:jz_bx}b. This demonstrates the fact that the plasma responds diamagnetically to the applied magnetic field. The enhanced magnetic field in the outer regions leads to stronger focusing of hot electrons which results in a more efficient sheath field in the presence of an external magnetic field (Fig. \ref{fig:sheath}b) compared with those without a magnetic field (Fig. \ref{fig:sheath}a).

The line out of the sheath field in Fig. \ref{fig:sheath}c shows that the magnitude of the sheath fields is comparable to that of without the magnetic field but it can enhance the proton energy by $\sim 30\%$($\sim$5.6 MeV). This energy enhancement is much less compared to that achieved by Arefive et.al \cite{arefiev2016enhanced}  $\sim 50\%$ ($\sim$9.5 MeV). This difference is because of a long pre-plasma considered in their simulations and the step density profile chosen in our case. When we use 10 $\mu$m long, exponential pre-plasma, the proton energy gets enhanced by $\sim 33\%$ ($\sim$ 9.8 MeV).

\begin{figure}
	\includegraphics[width=1\textwidth]{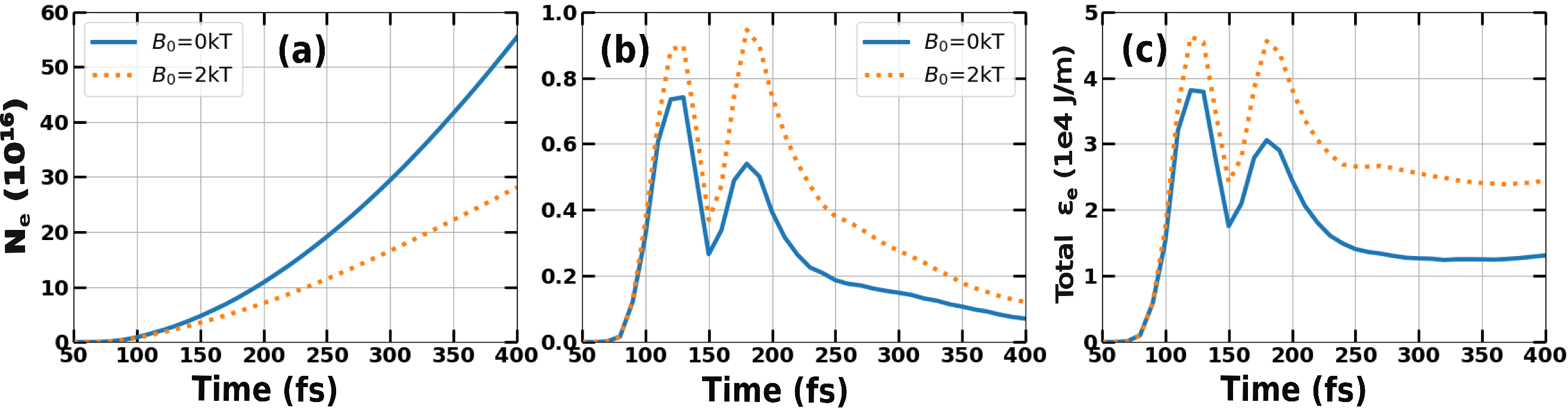}
 	\caption{(a) Total number of electrons, (b) electrons having energy greater than 1MeV are moving at the rear side of the target with time. (c) The total energy is carried by the total number of electrons at the rear side.}
  \label{fig:electron_number}
\end{figure}

To investigate the above-unexpected behavior of the sheath field at the rear side, the number of hot electrons moving to the rear side with time i.e. the electrons that are responsible for the sheath formation is plotted with time in Fig. \ref{fig:electron_number}a. It can be seen that in the presence of an external magnetic field, the number of electrons moving to the rear side gets reduced. But if we restrict ourselves to the electrons having energy greater than 1 MeV (Fig. \ref{fig:electron_number}b), we see that initially there is no difference in the number of electrons. After some time, the number of energetic electrons (energy $\geq$ 1 MeV) is more than without the magnetic field case. The same behavior is also followed by the total energy gained by these electrons (Fig. \ref{fig:electron_number}c).

Here, three points are worth noting. The first point is that there are two peaks and one local minimum in the number of energetic electrons and total energy carried by the electrons at the rear side. As the laser interacts with the target it transfers energy to the electrons and these energetic electrons start coming out at the rear side. Hence the energetic electron number and their energy continuously increase. At time 120 fs, the laser just leaves the target and electrons start oscillating i.e. they are pulled back into the target and their number at the rear side starts decreasing.  Around t=150 fs, electrons come out at the front side of the target and therefore their number is maximum at the front side and minimum at the rear side. After t=150 fs, the electrons again start moving towards the rear side and hence form the second peak of energetic electrons as shown in Figs. \ref{fig:electron_number}b and \ref{fig:electron_number}c. 

\begin{figure}
	\includegraphics[width=0.8\textwidth]{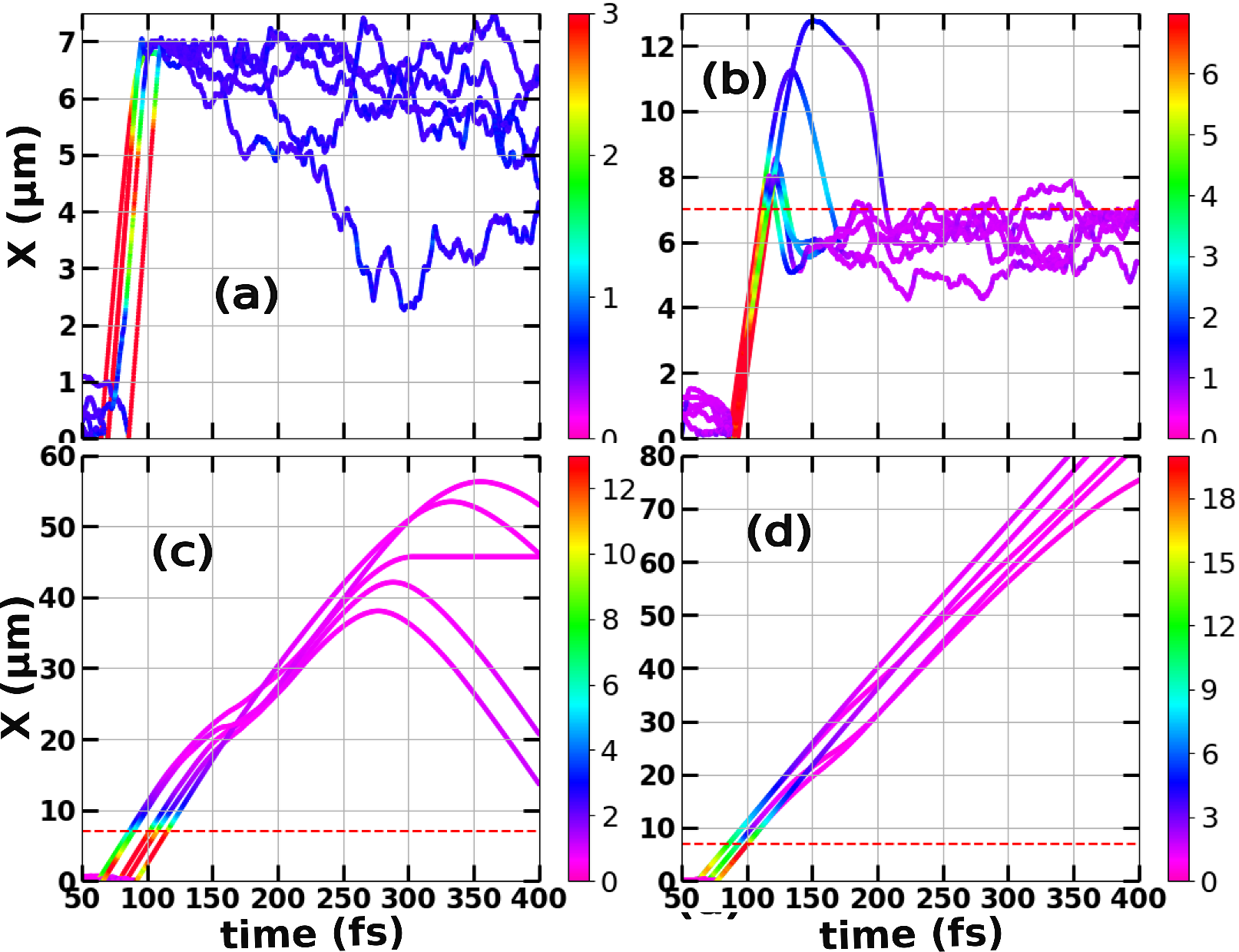}
 	\caption{Electrons longitudinal trajectory along the longitudinal direction with time and the color bar represents energy in MeV. There are four types of electrons depending on the initial energy gained from the laser a) the electrons that get trapped inside the target, b) the electrons that escape from the target but return after losing their energy to ions/protons, c) the electrons that escape from the target and get trapped inside the sheath, and d) the electrons that are energetic enough to even escape the sheath field and are lost from the simulation box. }
  \label{fig:c_trajectory}
\end{figure}

Secondly, the applied magnetic field not only restricts the electrons in the transverse direction but also restricts the low-energy electrons from coming out at the rear side of the target. On tracking the trajectories of a large number of electrons it is observed that there are typically four varieties of electron populations as shown in four subplots of Fig. \ref{fig:c_trajectory}. The low energetic electrons (Fig. \ref{fig:c_trajectory}a) gain energy up to 3 MeV at the front side of the target and remain trapped at the rear side within the target (x=7 $\mu$m) whereas the highly energetic electron(Fig. \ref{fig:c_trajectory}d), having initial energy of approximately 20 MeV, exit from the target rear side and escape from the simulation box. Both of these electron populations do not contribute to the sheath formation. The electrons that gain initial energy of $\sim$ 5-7 MeV exit from the target but are then pulled back to the target after losing their energy to the ions/protons (Fig. \ref{fig:c_trajectory}b).  The fourth kind of electrons with energy $\sim$ 10-12 MeV exit from the target rear and do not return to the target but get trapped in the sheath (Fig. \ref{fig:c_trajectory}c). 

\begin{figure}
	\includegraphics[width=.8\textwidth]{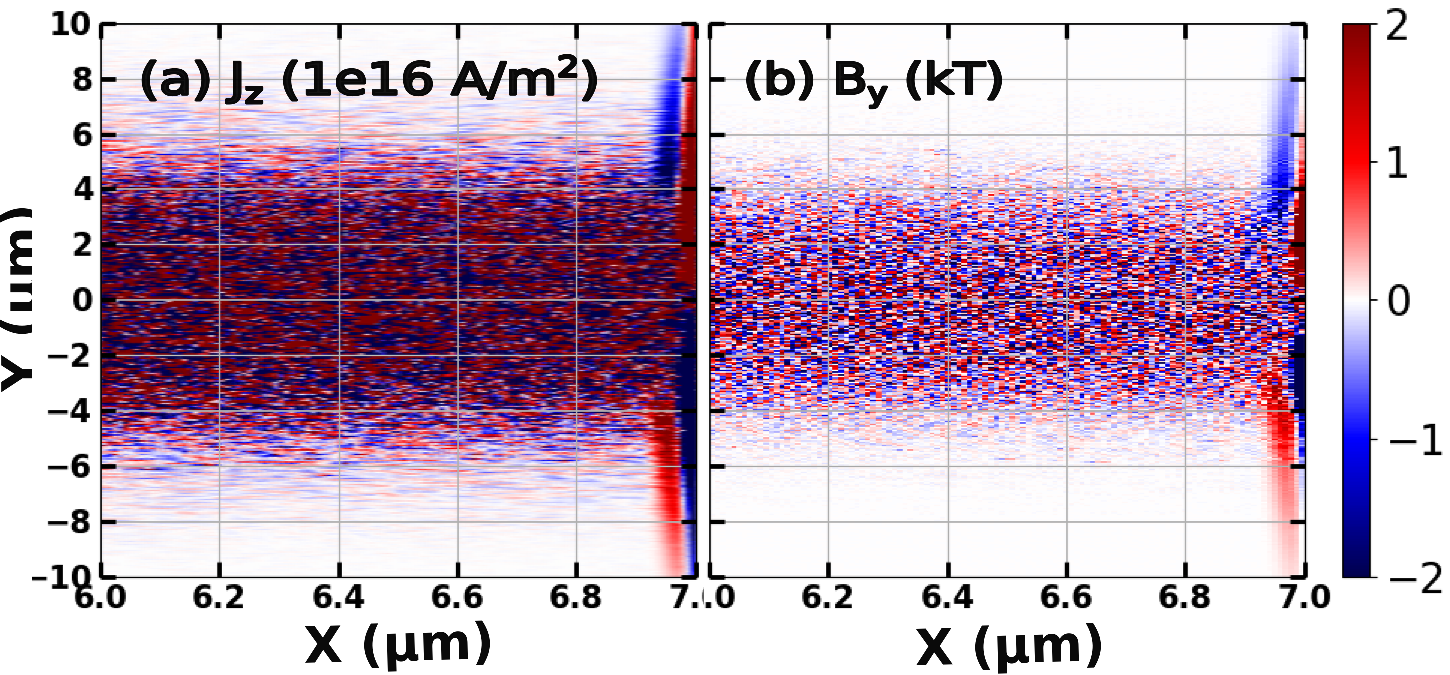}
 	\caption{The current density (a) and the magnetic field (b) generated inside the target, at time t=100 fs, for a single normal laser pulse in the presence of the external magnetic field.}
  \label{fig:jz_by}
\end{figure}

\begin{figure}
	\includegraphics[width=0.5\textwidth]{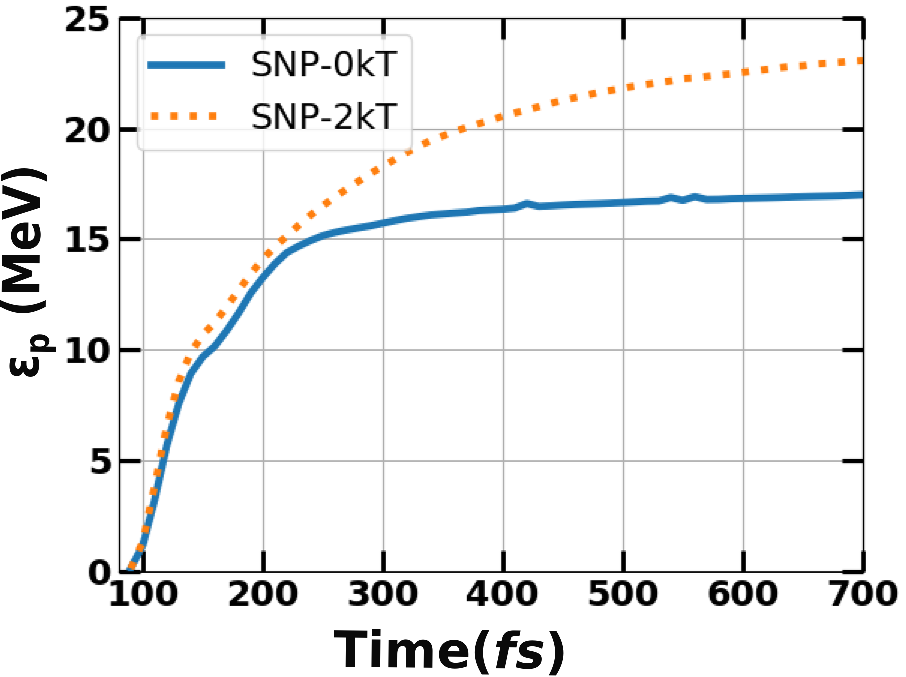}
 	\caption{The evolution of the proton cutoff energy with time, for SNP configuration.}
  \label{fig:cutoff_energy}
\end{figure}

The low energy electrons (Fig. \ref{fig:c_trajectory}a) trying to escape to the rear side of the target experience the induced magnetic field ($\bold{B_i}$) in the transverse direction just outside the target. Considering the upper half of the target (+y-axis), these electrons experience the force ($V_x\times B_i$) in -z-direction that provides them a circular path within a few nano-meters inside the rear side of the target. The y-component of electron velocity makes it a helical path. In 2D, it is not possible to see these helical trajectories of electrons but two current sheets can still be seen near the rear side of the target: one with current in +y direction (just inside the rear surface), and another with current in -y direction (a few nano-meters inside the rear surface). These current sheets are infinite in the z-direction, therefore the magnetic field($\bold{B_h}$) associated with these sheets at any point is independent of the distance from the sheet. The magnetic fields due to these two current sheets, of opposite polarity, add up in between them and cancel out outside. The current sheets and magnetic field ($\bold{B_h}$) can be seen in Figs. \ref{fig:jz_by}a and \ref{fig:jz_by}b, respectively.

The third interesting observation is that in the presence of a magnetic field, the overall energy carried by the electrons (Fig. \ref{fig:electron_number}c) is large even though the total number of electrons is lower (Fig. \ref{fig:electron_number}a). The enhancement in electrons' total energy in the presence of an external magnetic field can be explained with the help of ponderomotive force. The ponderomotive force experienced by the electrons /ions with and without the presence of the longitudinal external magnetic field\cite{goswami2021ponderomotive} is 

$\left( \frac{\partial u}{\partial t}\right) _{2kT} = - \frac{q^2_{i} \bigtriangledown |E|^2}{2m^{2}_{i} \omega} \frac{1}{\left( 1-\frac{\omega ^{2}_{ci}}{\omega^2 } \right)} $ 	and 	$\left( \frac{\partial u}{\partial t}\right) _{0kT} = - \frac{q^2_{i} \bigtriangledown |E|^2}{2m^{2}_{i} \omega} $

Where $q_i$, $\omega$, $\omega_{ci}$, $m_i$ are the charge, laser frequency, particle gyration frequency, mass of the particle, and $i$ is used to specify the type of particle i.e. electron or proton. As $\omega > \omega_{ci}$, the force experienced by the electrons or ions in the presence of a magnetic field is higher and hence they gain higher energies.

Moreover, it can be seen from Fig. \ref{fig:cutoff_energy} that initially (up to t=200 fs) there is no difference in the proton cutoff energies in the cases of with and without the external magnetic field. The difference starts appearing after this time as the cutoff energy tends to saturate beyond this point in the absence of magnetic field whereas the cutoff energy keeps increasing when the longitudinal magnetic field is present. 

\begin{figure}
	\includegraphics[width=.8\textwidth]{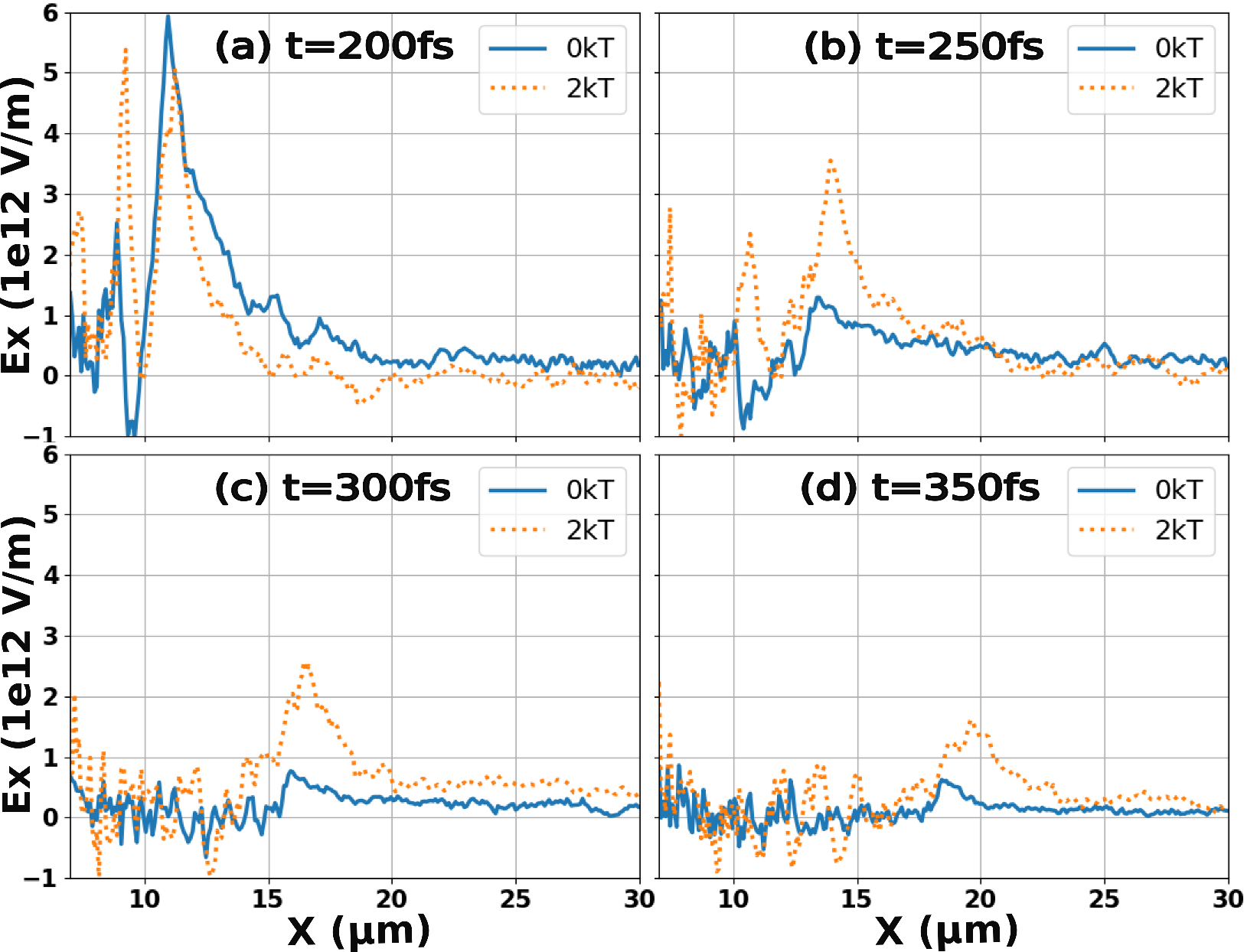}
 	\caption{Line out of the sheath electric field at the rear side of the target at time (a) 200 fs, (b) 250 fs, (c) 300 fs, and (d)350 fs.}
  \label{fig:c_Ex_lineout}
\end{figure}

To understand this observation we plot, in Fig. \ref{fig:c_Ex_lineout}, the lineout of the longitudinal sheath field along the horizontal symmetry axis of the target. It is found that the sheath field in the presence of an external magnetic field is maintained comparatively for a longer time. It can be noted from the subplots of Fig. \ref{fig:c_Ex_lineout} at t=250 fs, 300 fs, and 350 fs, that the sheath field in the absence of the magnetic field has decayed whereas there still exists sufficiently strong sheath field in the presence of external magnetic field. Hence, after approximately 200 fs, the cutoff energy becomes saturated in the absence of the magnetic field while it continuously increases and tends to saturate after a longer duration in the presence of the external magnetic field.

\section{Oblique incidence in the presence of B}
\begin{figure}
	\includegraphics[width=1\textwidth]{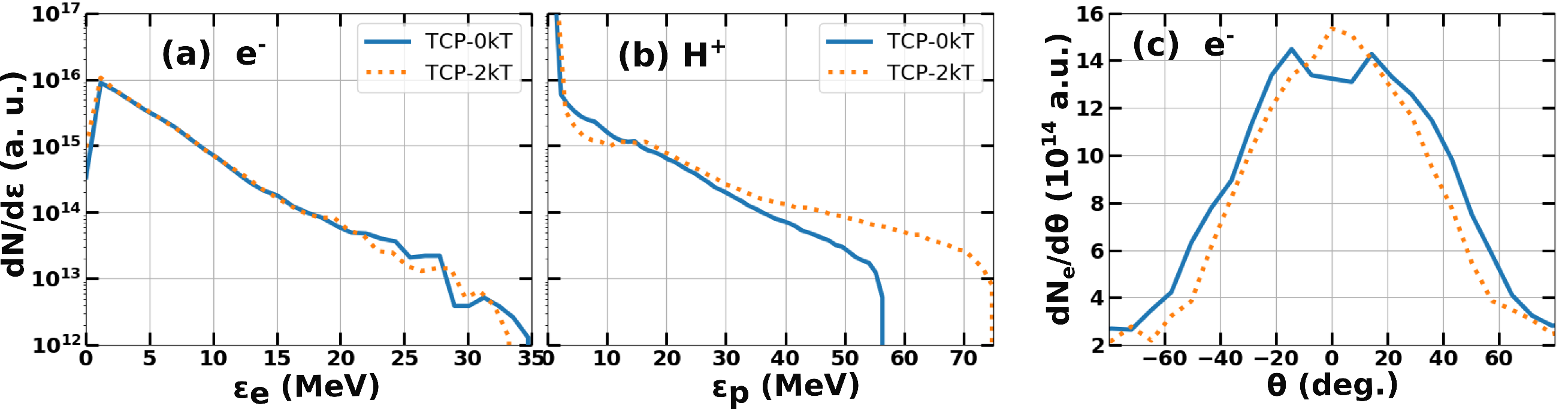}
 	\caption{(a) Electron energy distribution, (b) protons energy distribution, and (c) electron angular distribution without (solid line) and with (dotted line) an external magnetic field for TCP. The electron energy distribution and angular distributions are shown at t=100 fs whereas the energy distribution of the proton is recorded at t=870 fs.}
  \label{fig:c_dist}
\end{figure}

Now the effect of an external magnetic field is investigated in the case of target irradiation by two oblique laser pulses (two colliding pulses or TCP configuration). It is observed that the electron energy spectra are slightly affected by the externally applied magnetic field, as shown in Fig. \ref{fig:c_dist}a, but the proton energy spectra exhibit a significant amplification in the cut-off energy, see Fig. \ref{fig:c_dist}b. This is similar to the observations with SNP discussed in the previous section and also reported previously in  Ref. \cite{arefiev2016enhanced, weichman2020generation}. 

\begin{figure}
	\includegraphics[width=.5\textwidth]{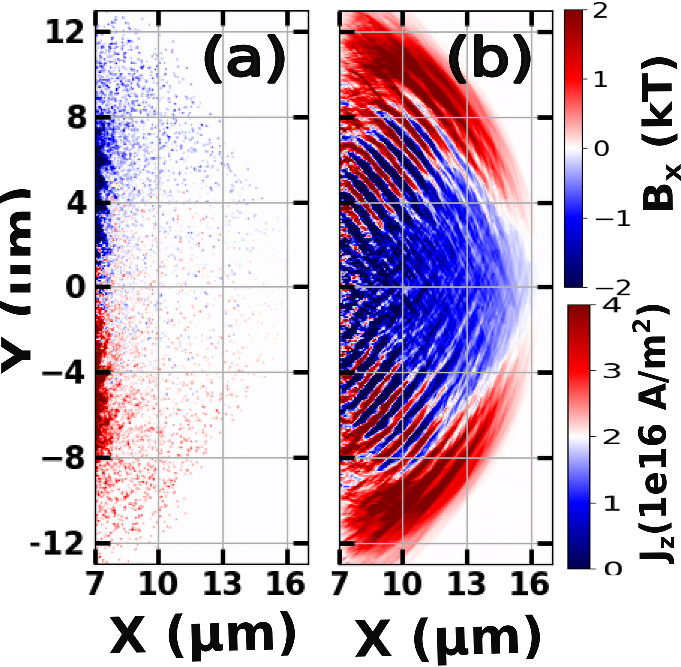}
 	\caption{Induced current density along the z-direction, $j_z$ (left), and modified longitudinal magnetic field (right) at the rear side of the target for TCP at time t=100 fs, in the presence of an external magnetic field.}
  \label{fig:c_jz_by}
\end{figure}

\begin{figure}
	\includegraphics[width=.6\textwidth]{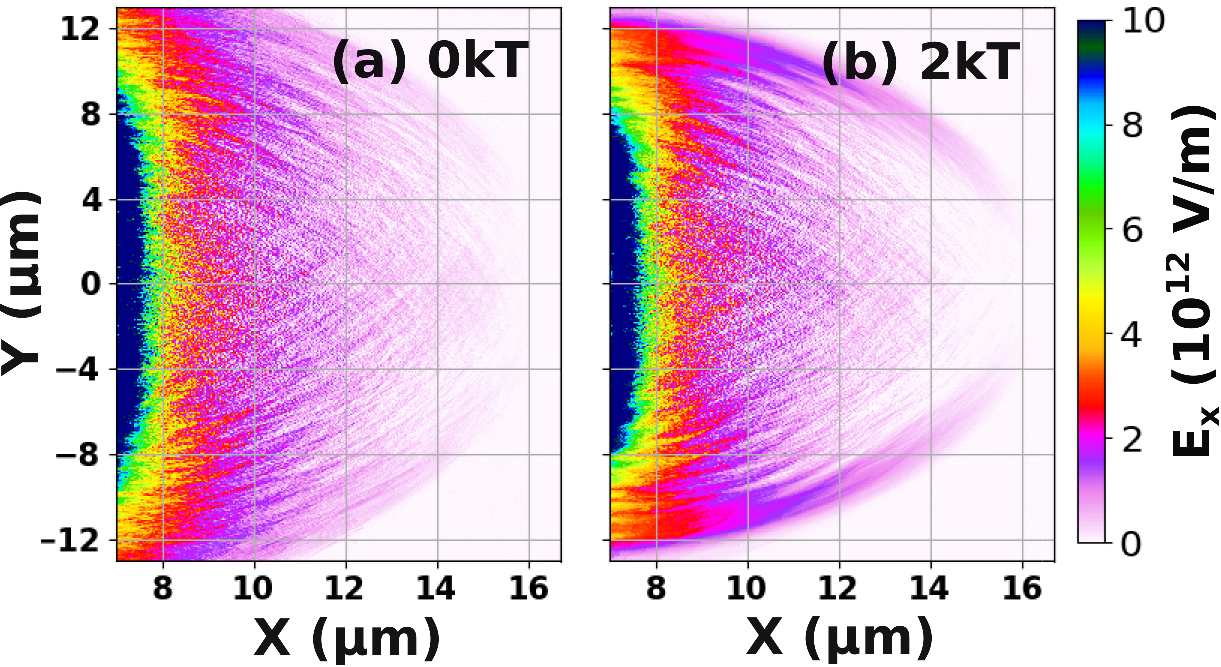}
 	\caption{The longitudinal electric field at the rear side of the target without (left) and with (right) an external magnetic field for TCP at time t=100 fs.}
  \label{fig:c_Ex}
\end{figure}

 \begin{figure}
	\includegraphics[width=1\textwidth]{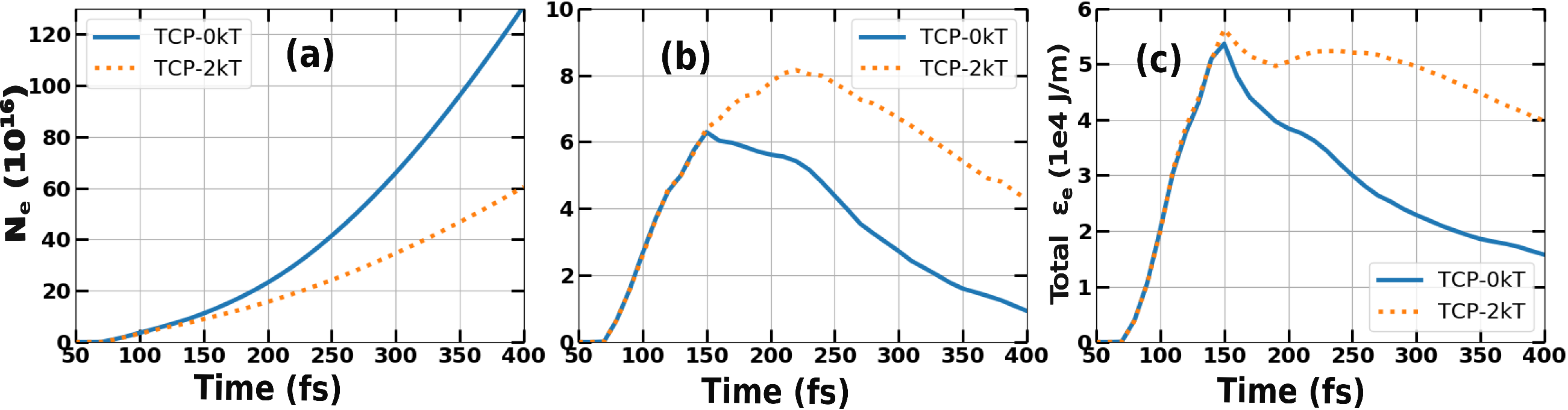}
 	\caption{(a) Total number of electrons, (b) electrons having energy greater than 1 MeV are moving at the rear side of the target with time for TCP. (c) The total energy is carried by the total number of electrons at the rear side.}
  \label{fig:c_dn}
\end{figure}

Now, even for the single oblique incidence laser (SOP) and the step density flat target, the energy absorption by the electrons is enhanced as the angle of incidence is raised from 0 degrees (normal incidence) to 45 degrees. This can be attributed to the involvement of vacuum heating\cite{brunel1987not,yogo2015ion} along with the $\bold{J\times B}$\cite{mulser2008collisionless,kruer1985j} mechanism. In the TCP case, there is a further enhancement in the electron cutoff energy because the two laser pulses (of half intensity) interfere in front of the target and the maximum resultant field interacting with the target becomes $\sqrt{2}$ times that of a single pulse\cite{ferri2019enhanced, rahman2021particle}. An additional benefit of the TCP is its ability to eliminate the DC current that is established along the surface. This DC current in the SOP configuration produces an extra magnetic field that hinders the fast electrons from returning to the target by deflecting them\cite{gibbon2005short} thus inhibiting the hot electron generation inside the target.

Figure \ref{fig:c_jz_by} shows that in the TCP configuration, the current generated due to plasma plume expansion (Fig. \ref{fig:c_jz_by}a) and the modified longitudinal magnetic field (Fig. \ref{fig:c_jz_by}b) have similar pattern as in the case of SNP configuration but their magnitudes are enhanced by factors of 1.7 and 2.4, respectively. The sheath field at the rear side also has a similar pattern as that of SNP i.e. focused with a magnetic field (Fig. \ref{fig:c_Ex}b) than that without the magnetic field (Fig. \ref{fig:c_Ex}a). In the presence of the external magnetic field ($\sim$2 kT), the enhancement in the maximum sheath field magnitude for the TCP configuration is 2.9 times its maximum magnitude in the SNP configuration. For comparison, in the absence of any external magnetic field, the sheath field magnitude in the TCP configuration is found to be approx. 2.2 times larger as compared to the SNP configuration.

In the TCP configuration (Fig. \ref{fig:c_dn}a), the total number of electrons moving out of the rear surface of the target is reduced in the presence of the externally applied magnetic field as is also observed in the case of SNP configuration (Fig. \ref{fig:electron_number}a). However, if we count the electrons having energy greater than 1 MeV, their number is higher in the presence of the external magnetic field than without it (Fig. \ref{fig:c_dn}b). This behavior is also similar to our numerical observations in the SNP case (Fig. \ref{fig:electron_number}b).

 Another point to note is that in the case of SNP configuration, the number of electrons and their total energy starts decreasing around t=130 fs, and the minimum is reached at t=150 fs. On the other hand, in TCP configuration, this decrease starts at t=150 fs. This slight delay in the TCP configuration is because of the extra distance traveled by the laser pulse, in the oblique incidence case, before interacting with the target. Also, the minima observed in the time evolution of the number of electrons at the rear side as well as their energy content, for the case of SNP configuration, is absent in the TCP configuration which possibly can be attributed to the delay in the laser pulse reflection from the target in the TCP case.

\section{Conclusions}
In this study, using two-dimensional particle-in-cell simulations, the effect of an externally applied kilo-Tesla magnetic field on the spectra of electrons as well as of protons has been thoroughly investigated in the cases of a single normally incident laser pulse (SNP), a single obliquely incident laser pulse (SOP), and two oblique colliding laser pulses (TCP). 

For SNP, the proton cutoff energy is shown to be increased due to an external kilo-Tesla magnetic field from 19 MeV to 24.62 MeV ( approx. 30 \% enhancement). On the other hand, in the single oblique pulse (SOP) configuration, the proton cut-off energy is found to be enhanced by the external magnetic field, from 38.4 MeV to 55.2 MeV (approx. 44\% enhancement). Finally, in the oblique TCP configuration, the presence of an external magnetic field enhances the proton cut-off energy from 56.9 MeV to 75.5 MeV ( approx. 33\% enhancement). Thus the TCP configuration produces the highest cutoff energy when an external magnetic field, of kilo-Tesla level, is present. 

From the above, it can be concluded that while a single oblique pulse without an external magnetic field is more effective than a single normal pulse in the presence of an external kilo-Tesla magnetic field, two colliding pulses (each with a halved intensity compared to SNP/SOP) without a magnetic field are found to be better than the single oblique pulse even with an external magnetic field. The combination of two colliding pulses (TCP) configuration along with an external kilo-Tesla magnetic field is found to be the most effective arrangement for producing energetic protons.

\begin{acknowledgments}
The authors would like to acknowledge the EPOCH consortium, for providing access to the EPOCH-4.9.0 framework \cite{arber2015contemporary}, and high-performance computing (HPC) facility at the Indian Institute of Technology Delhi for computational resources. IK also acknowledges the University Grants Commission (UGC), govt. of India, for his senior research fellowship (Grant no. 1306/(CSIR-UGC NET DEC. 2018)).  
\end{acknowledgments}
\section*{Data Availability}
The data that support the findings of this study are available from the authors upon reasonable request. 
%
\end{document}